# Ultrathin broadband reflective optical limiter

Chenghao Wan[1,2], Zhen Zhang[3], Jad Salman[1], Jonathan King[1], Yuzhe Xiao[1], Zhaoning Yu[1,4], Alireza Shahsafi[1], Raymond Wambold[1], Shriram Ramanathan[3], and Mikhail A. Kats[1,2,4*]

[1] *Department of Electrical and Computer Engineering, University of Wisconsin-Madison, Madison, Wisconsin 53706, USA*
[2] *Department of Materials Science & Engineering, University of Wisconsin-Madison, Madison, Wisconsin 53706, USA*
[3] *School of Materials Engineering, Purdue University, West Lafayette, IN 47907, USA*
[4] *Department of Physics, University of Wisconsin-Madison, Madison, Wisconsin 53706, USA*

*\*Email: mkats@wisc.edu*

# Abstract

Optical limiters are nonlinear devices that feature decreasing transmittance with increasing incident optical intensity, and thus can protect sensitive components from high-intensity illumination. The ideal optical limiter reflects rather than absorbs light in its active ("limiting") state, minimizing risk of damage to the limiter itself. Previous efforts to realize reflective limiters were based on embedding nonlinear layers into relatively thick multilayer photonic structures, resulting in substantial fabrication complexity, reduced speed and, in some instances, limited working bandwidth. We overcome these tradeoffs by using the insulator-to-metal transition in vanadium dioxide ($VO_2$) to achieve intensity-dependent modulation of resonant transmission through aperture antennas. Due to the dramatic change of optical properties across the insulator-to-metal transition, low-quality-factor resonators were sufficient to achieve high on-off ratios in device transmittance. As a result, our ultra-thin reflective limiter (thickness ~1/100 of the free-space wavelength) is broadband in terms of operating wavelength (> 2 µm at 10 µm) and angle of incidence (up to ~50° away from the normal).

# Introduction

The growing ubiquity of high-power light sources[1,2] and the increasing sensitivity of photodetectors and cameras[3,4] motivate designs of optical limiters that can protect delicate optical components and human eyes against high-intensity illumination. Optical limiters are designed to be transparent for low-intensity incident light (open state), but should have gradually decreasing transmittance for increasing intensity (limiting state) such that the output intensity remains roughly constant[5,6,7,8,9].

Existing optical limiters are primarily based on nonlinear absorption[9,10], with alternatives including nonlinear refraction[11] or nonlinear scattering[12,13]; however, each of these mechanisms has drawbacks. The use of nonlinear refraction or scattering requires additional care to redirect the refracted or scattered light away from any light-sensitive devices, so nonlinear absorption is typically preferred. A variety of nonlinear absorption mechanisms across many material systems have been used, including reverse saturable absorption in organic/organometallic polymers[8], multi-photon absorption in liquid crystals[14] and transition metal dichalcogenides[15], free-carrier absorption in semiconductors[16], and many others[6,9,17]. Despite their popularity, absorptive limiters are often not ideal because absorption of intense incident light can damage the limiter itself, e.g., via melting, sputtering, delamination, or ablation.

The ideal solution is thus a limiter that transitions from a high-transmittance state at low intensity to a reflecting state at high intensity. Such devices have recently been proposed and demonstrated using multilayer photonic structures with the nonlinear medium comprising gallium arsenide (GaAs)[18] or germanium-antimony-tellurium (GST)[19]. In the first demonstration based on GaAs[18], the spectral and angular bandwidths were small due to the large quality factor of the photonic-crystal resonator, which was needed to enhance the optical nonlinearity. In the recent demonstration based on GST[19], a complex multilayer structure with multiple embedded GST layers was used to achieve broadband performance, enabled by the large effective nonlinearity due to the phase change of GST. Note, however, that the use of multiple active layers within a thick structure can limit the switching speed, and the nonvolatile phase change in GST means that a limiter based on this material can only switch to the limiting state once before it must be manually "reset".

In this paper, we demonstrate an ultrathin, volatile, reflective optical limiter that only comprises two subwavelength-thickness functional layers and operates over a broad range of wavelengths and incidence angles. This is achieved using resonant transmission through a metallic frequency-selective surface (FSS) comprising low-quality-factor aperture antennas, made optically responsive using the insulator-to-metal transition (IMT) of vanadium dioxide ($VO_2$), which can be triggered optically via a photothermal process[20,21]. The very large change in optical properties across the IMT[22] enables the strong modulation of resonant transmission through the aperture antennas, resulting in a limiter with broadband transmission in the open state and low transmission and absorption in the limiting state.

## Results

Our design comprises a gold FSS, a thin film of $VO_2$, and a transparent substrate (GaAs), as shown in Fig. 1(a). The IMT in $VO_2$ can be thermally triggered by heating to ~70 °C[23,24], and can be the source of photothermal nonlinearity[22,25] that is orders of magnitude stronger than conventional nonlinearities (e.g., the Kerr effect[26,27]). Indeed, the limiting characteristics of $VO_2$ films based on this IMT are well-known [21,28,29]. However, metal-state $VO_2$ is not nearly as reflective as noble metals (e.g., gold and silver)[22,30], and therefore a nontrivial amount of incident power can still be absorbed by the metal-state $VO_2$.

Our use of an FSS on top of a $VO_2$ film significantly reduces the limiting-state absorption, enabling the use of the IMT in $VO_2$ for reflective limiting. The gold FSS in our design is an array of close-packed cross-slit aperture antennas within a thin film of gold [Fig. 1(a)] that has high resonant transmittance when the nearby $VO_2$ film is in the insulating phase. Once the IMT is triggered by sufficiently intense incident light, the resonant transmittance is expected to reduce dramatically via two mechanisms. First, the resonance is frequency-shifted by the large change in the real part of the refractive index of $VO_2$. Second, the increasing loss in the $VO_2$ further suppresses the amplitude of the resonant transmission. These two mechanisms result in a large reduction of the device transmittance as the $VO_2$ undergoes the IMT. For the substrate, we chose undoped GaAs (001) since it is mostly transparent across the mid infrared[31]. We used finite-difference time-domain (FDTD) simulations (implemented in Lumerical FDTD) for design and optimization.

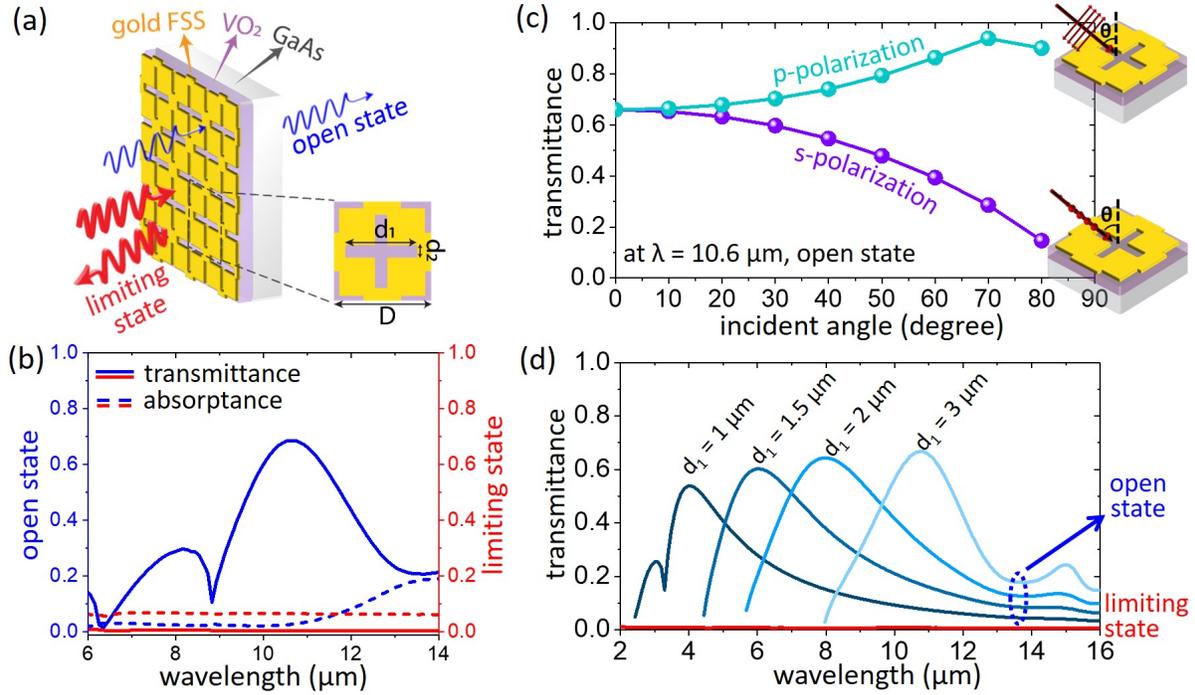

**Figure 1**. **(a)** Schematic of an optical limiter based on a gold FSS on top of a thin film of $VO_2$, with GaAs (001) as the transparent substrate. **(b)** Simulated open- and limiting-state transmittance and absorptance of the structure in "(a)" with $d_1$ = 0.2 μm, $d_2$ = 3.1 μm and D = 3.5 μm. The thickness of gold and $VO_2$ are 50 nm and 100 nm, respectively. The central wavelength of the open-state transmittance peak is λ = 10.6 μm, with peak transmittance of 0.7. The limiting-state transmittance is smaller than 0.01, accompanied by absorptance of ~0.06 at all wavelengths. **(c)** Simulated angle-dependent open-state transmittance of the design in "b" for both s- and p-polarizations at λ = 10.6 μm. **(d)** The central wavelength of the open-state resonant transmittance is tunable between 4 and 11 μm by changing $d_1$ from 1 to 3 μm. The value of $d_2$ was fixed at 0.2 μm. The value of D is always set to ($d_1$ + 0.4) μm.

The structure was optimized for maximum transmittance at a target wavelength when in the open state, and maximum reflectance when in the limiting state. Specifically, we considered three figures of merit: a) high open-state transmittance that ensures high transmission efficiency of low-intensity light; b) low limiting-state transmittance that ensures effective limiting of high-intensity light; and c) low limiting-state absorptance that enables a high damage threshold. For this paper, we chose λ = 10.6 μm as the central wavelength of the open-state resonant transmittance. Using the temperature-dependent refractive indices of $VO_2$ that were characterized in our previous work (ref. 22), we optimized the structural parameters of the aperture antennas, including length ($d_1$) and width ($d_2$) of the aperture antenna and the periodicity (D). Note that there are tradeoffs between our three figures of merit. For example, a larger $d_2$ or a smaller D helps to improve the open-state transmittance but also increases the limiting-state absorptance (see detailed discussion in *Supplementary Information 1*).

After considering these tradeoffs during optimization, we settled on $d_1 = 3.1$ μm, $d_2 = 0.2$ μm, and $D = 3.5$ μm. The thicknesses of gold or $VO_2$ have much less influence on our figures of merit because they are much smaller than the wavelength (thickness of ~100 nm ≪ operational wavelength of 10.6 μm). Therefore, we chose the thicknesses of gold and $VO_2$ to be 50 nm and 100 nm, respectively. The resulting structure features a broad transmittance band (FWHM > 2 μm) centered at λ = 10.6 μm in the open state and close-to-zero transmittance (~0.008) in the limiting state [Fig. 1(b)]. The limiting-state absorptance (~0.06 at all wavelengths) is significantly reduced compared to that of the bare $VO_2$ film with no FSS (~0.2, see details in *Supplementary Information 2*), thus enhancing the damage threshold of the device. The limiting-state absorptance can be further decreased by reducing the area density of the aperture antennas; however, this comes at the cost of the amplitude and bandwidth of the open-state transmittance peak (*Supplemental Information 1*). Due to the subwavelength thickness of the FSS-$VO_2$ layer and the symmetry of the aperture antennas, the open-state transmittance remains high for oblique incidence up to ~50° from the normal for both s- and p-polarizations [Fig. 1(c)]. Because insulating thin-film $VO_2$ has low optical loss in the infrared from 2 to 11 μm[22], the transmittance peak of the limiter can be shifted to anywhere within this window by changing the length of aperture antennas [Fig. 1(d)].

Our fabrication process includes four steps: $VO_2$ synthesis, e-beam lithography, gold evaporation, and lift-off [Fig. 2(a)]. First, $VO_2$ was magnetron-sputtered onto double-side-polished undoped GaAs (001) wafers (see details in *Materials and Methods*). The thickness of the resulting film is (105 ± 6) nm, measured by scanning electron microscopy (SEM) imaging of the cross section and atomic force microscopy (AFM) imaging of the surface. The stoichiometry of the film was confirmed by Raman spectroscopy measurements at 30 °C and 100 °C (see details in *Supplementary Information 3*). The IMT of the film was confirmed by temperature-dependent near-normal-incidence transmission/reflection measurements using a Fourier-transform spectrometer (FTS) [Fig. 2(b)]. All spectra were collected at temperatures between 30 °C and 100 °C (first heating, then cooling) with steps of 2 °C and a ramping rate of 1 °C/min. The as-grown film featured an IMT from ~70 to 82 °C when heated [Fig. 2(c, d)].

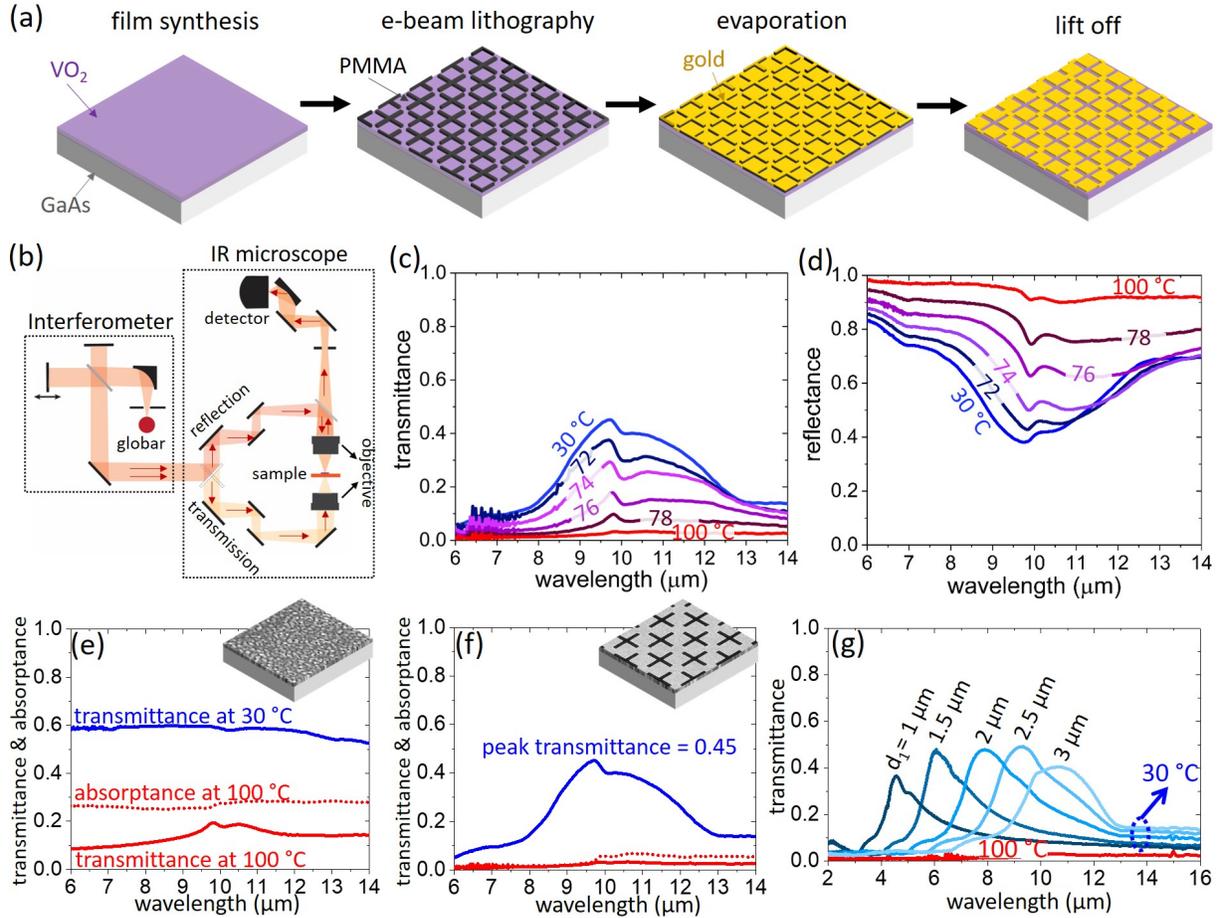

**Figure 2. (a)** Fabrication flow: we first synthesized ~100-nm $VO_2$ on GaAs wafer using magnetron sputtering. The as-grown film went through an e-beam lithography and development process, resulting in an array of cross-shape blocks of PMMA. Then, we evaporated 50 nm of gold on top, and performed a lift-off process, leaving an array of cross-slit apertures within the gold film. **(b)** A schematic of simplified optical path of our FTIR transmission and reflection measurements, using an infrared microscope (Bruker Hyperion 2000) attached to a Fourier-transform spectrometer (Bruker Vertex 70). **(c, d)** Transmittance and reflectance measurements of our fabricated FSS-$VO_2$ limiter when $VO_2$ in its pure insulating phase (30 °C), pure metallic phase (100 °C), and intermediate phases across the IMT (at 72, 74, 76, and 78 °C). **(e)** Transmittance and absorptance spectra of the as-grown $VO_2$ film, for both insulating (30 °C) and metallic (100 °C) phases. **(f)** Transmittance and absorptance spectra of the FSS-$VO_2$ limiter at 30 °C (open state) and 100 °C (limiting state). **(g)** Measured transmittance of the FSS-$VO_2$ limiters with aperture lengths $d_1$ of 1, 1.5, 2, 2.5, and 3 μm, for both open (30 °C) and limiting (100 °C) states.

The as-grown $VO_2$ film features broadband transmittance of ~0.6 in the insulating phase and relatively low transmittance when thermally biased to the metallic state [Fig. 2(e)]. The features at ~10 μm of the metallic-state $VO_2$ are possibly due to a very thin surface oxide of $V_2O_5$ on the as-grown film[22,32] (more details in Supplementary Information 3). We note that, as shown later, such thin surface oxide does not substatially affect the performance of our fabricated limiter. The absorptance spectrum at 100 °C (absorptance = 1 – transmittance – reflectance) indicates that the metallic-state $VO_2$ has a relatively high absorptance of ~0.3 at all wavelengths.

Then, we built a 50-nm gold FSS on top of the as-grown $VO_2$ via steps as shown in Fig. 2(a) (see details in *Materials and Methods*). At 30 °C (device in the open state), the fabricated FSS-$VO_2$ limiter features a peak transmittance of 0.45 at λ = 9.8 μm, while the transmittance decreased to ~0.03 when the temperature was increased to 100 °C (device in the limiting state) [Fig. 2(f)]. The measured reflectance in the limiting state is >0.9 at all wavelengths [Fig. 2(d)], indicating that the limiting-state absorptance is <0.06 [Fig. 2(f)], which is substantially reduced compared to that of the bare $VO_2$ film [Fig. 2(e)]. To demonstrate that our design is feasible for wavelengths across the mid infrared, we also fabricated samples with different aperture lengths. The measurements [Fig. 2(g)] showed that our limiters with different aperture lengths ($d_1$ = 1, 1.5, 2, 2.5, and 3 μm) have central wavelengths of resonant transmittance ranging from 4 to 11 μm, as expected based on the simulations in Fig. 1(d).

The measured spectra of the fabricated FSS-$VO_2$ limiter [Fig. 2(f)] essentially agree with our simulations [Fig. 1(b)], though there are discrepancies in the amplitudes and peak positions of the open-state transmittance. The amplitude of experimental open-state transmittance is lower than that of the simulation, likely because we did not consider the backside of the substrate in the FDTD simulation. After taking it into account, the simulated peak value matches well with that of the experimental result (*Supplemental Information 4*). In practice, one can manufacture anti-reflection coatings on the backside of the GaAs substrate to achieve the performance similar to simulations. Also, the open-state central wavelength of the fabricated limiter is blue-shifted by ~0.8 μm compared to our simulation results primarily due to two experimental errors. First, the aperture length and width in the fabricated FSS are ~3 μm and ~0.17 μm, respectively (*Supplementary Information 4*), both smaller than those we used in simulations. Second, the refractive indices of $VO_2$ that we used for simulation were not extracted from the film that we used for fabrication. Small differences in refractive-index values are expected between $VO_2$ films, though they share similar tendencies in this infrared wavelength region[22].

Despite the shift of the transmittance peak, our fabricated limiter still features reasonable open-state transmittance of 0.36 at λ = 10.6 μm, enabling us to use a continuous-wave $CO_2$ laser to test our design. In our intensity-dependent transmission measurement setup [Fig. 3(a)], the sample was mounted on a heat stage with a through-hole aperture used to thermally bias the device, which was necessary since the maximum incident intensity achieved in our setup was not high enough to photothermally trigger the IMT from room temperature. The linearly polarized laser beam was first expanded using an infrared concave lens, then collimated (beam diameter ~10 mm) and focused onto the sample using two infrared convex lenses (focal length f = 10 cm), resulting in near-normal incidence (NA ~ 0.05) with a maximum incident intensity of ~6.5 kW/cm$^2$ (*Supplementary Information 5*). A rotating wire-grid infrared polarizer was used as a tunable attenuator, yielding an incident power range from less than 0.1 mW to ~190 mW. Two more convex lenses were used to re-focus light that passed through the sample onto a power meter (Thorlabs S132C). To avoid the thermal hysteresis of the $VO_2$[33] from influencing the measurements, we heated the sample from room temperature to a set bias temperature for each data point, then turned on the laser beam and recorded the transmitted power once the bias temperature was reached.

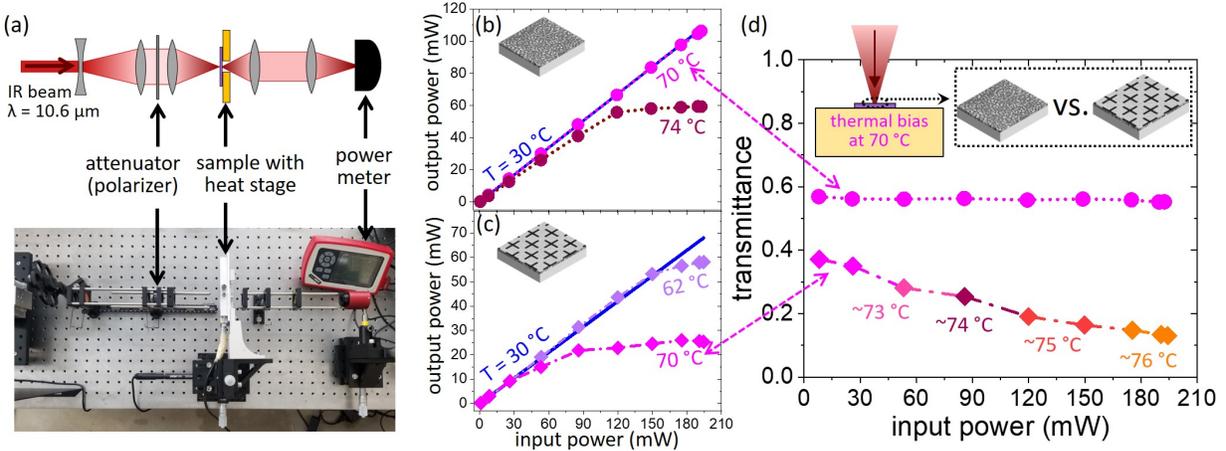

**Figure 3 (a)** Our measurement setup using a continuous-wave $CO_2$ laser (λ = 10.6 μm) as the light source and a heat stage for thermal biasing. **(b)** Power-dependent transmission measurements of the bare $VO_2$ film when it was thermally biased at T = 30, 70, and 74 °C. **(c)** Power-dependent transmission measurements of the FSS-$VO_2$ limiter when it was biased at T = 30, 62, and 70 °C. **(d)** A comparison between the power-dependent transmittance of the bare $VO_2$ film and the FSS-$VO_2$ limiter, when the heat stage was set at 70 °C. Each transmittance can be correlated with the local temperature at the beam spot using our previous temperature-dependent transmittance measurements, i.e., results in Fig. 2(c) and Fig. 4(b).

We measured the transmission of both the bare $VO_2$ film and fabricated FSS-$VO_2$ limiter for a series of incident powers and at different temperatures (i.e., temperature of the stage). For the bare $VO_2$ film [Fig. 3(b)], we did not observe nonlinear transmission (i.e., limiting behavior) when the temperature was set at or below 70 °C, which is at the edge of the IMT for increasing temperature [Fig. S3(c) in the *Supplementary Information*]. Limiting behavior became apparent when the film was thermally biased to 74 °C, which is already an intermediate temperature within the IMT. The throughput power was limited to ~55 mW at high-power incidence (i.e., incident power greater than 120 mW). In contrast, the FSS-$VO_2$ limiter featured obvious nonlinear transmission at ~70 °C for incident power starting at 30 mW [the pink curve in Fig. 3(c)]. The transmitted power was limited to ~25 mW for incident power from 90 to 190 mW. Furthermore, at a lower biasing temperature of 62 °C, which is far outside the IMT temperature range, we also observed limiting behavior when the incident power was increased above 150 mW. Such limiting behavior was not found in the bare $VO_2$ film using the same experimental conditions.

As expected from the initial simulations [Fig. 1(b)], the introduction of the FSS resulted in much lower limiting-state absorption and high transmittance contrast between the two states compared to the bare $VO_2$ film. Somewhat counterintuitively, the introduction of the FSS also resulted in limiting action starting at lower incident powers or lower bias temperature (i.e., a reduction of the limiting threshold). As shown in Fig. 3(d), when biased at 70 °C, the bare $VO_2$ film featured constant (i.e., linear) transmittance at all input powers, while the FSS-$VO_2$ limiter featured a decreasing (i.e., nonlinear) transmittance as the incident power increases, indicating that the limiter was photo-thermally heated to intermediate temperatures in the IMT region at the beam spot. For each transmittance, we can approximate the corresponding local temperature within the $VO_2$ film at the beam spot using our temperature-dependent transmittance measurements shown in Fig. 2(c).

The limiting threshold and required thermal bias depend strongly on the open-state absorptance of the device. In the case of bare thin-film $VO_2$, the open-state absorptance is close to zero [Fig. 4(a)], so a sufficiently high incident power is required to trigger the IMT. Alternatively, the film can be thermally biased part of the way into the IMT [e.g., at 74 °C, the dark red curve in Fig. 3(b)], where the absorptance has begun to rise even in the absence of incident light. The introduction of the FSS results in much-higher open-state absorptance of ~0.12 in our device [Fig. 4(a)], enabling limiting action that begins far outside the IMT [e.g., at 62 °C, the purple curve in Fig. 3(c)]. Furthermore, in the FSS-$VO_2$ limiter, both the higher open-state absorptance and the increase of absorptance at the onset of the IMT [Fig. 4(a)] result in faster turn-on compared to the bare $VO_2$ film. As the IMT evolves toward the pure metallic phase, the subsequent decrease of the absorptance starts to slow down the IMT, resulting in limiting performance for a broad intensity range [Fig. 3(c)].

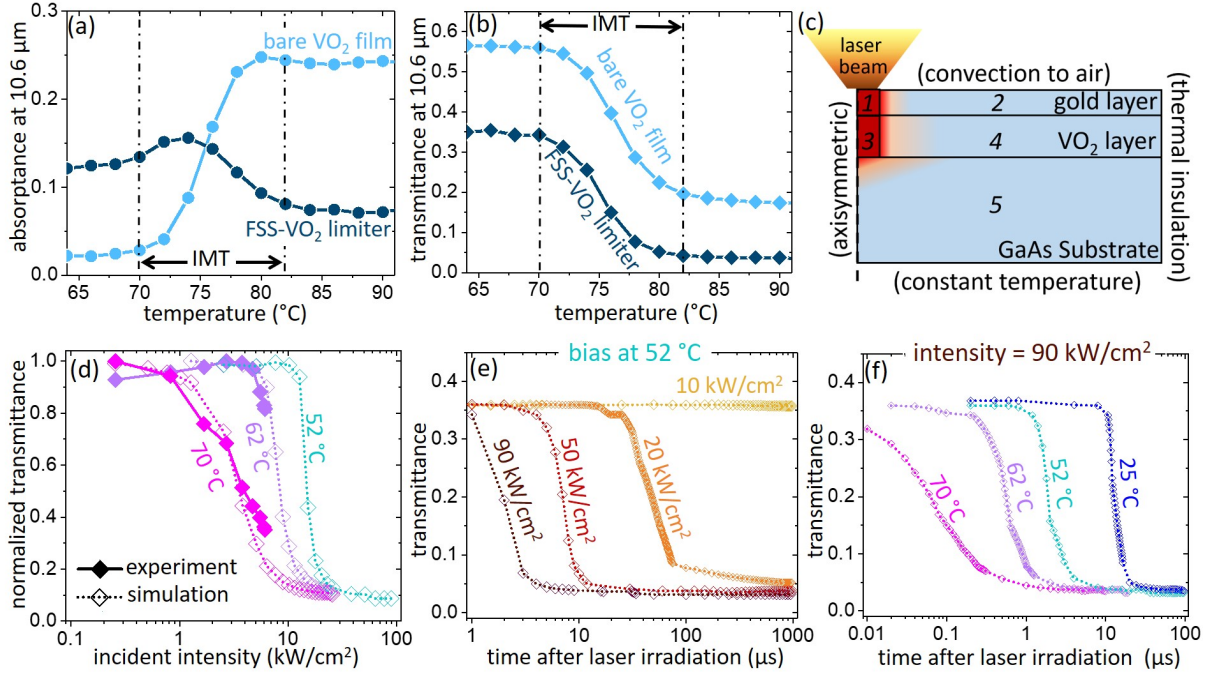

**Figure 4.** Measured temperature-dependent **(a)** absorptance and **(b)** transmittance of both the bare $VO_2$ film and FSS-$VO_2$ limiter at λ = 10.6 μm. **(c)** Geometry for photothermal simulations. The axisymmetric model consists of five domains of which 1 and 3 experience absorption due to the laser irradiation. **(d)** Comparison between the simulations and the experimental results in Fig. 3(c). **(e)** Simulated time-resolved transmittance of the FSS-$VO_2$ limiter for incident intensities of 10, 20, 50, and 90 kW/cm², when it is biased at 52 °C. **(f)** Simulated time-resolved transmittance of the FSS limiter for an incident intensity of 90 kW/cm², when it is thermally biased at 25, 52, 62, and 70 °C.

To quantitively understand our intensity-dependent measurements, we built a photothermal simulation using COMSOL Multiphysics. Our model consists of an axisymmetric geometry of five domains centered on the center axis of the laser beam [Fig. 4(c)]. The heat generation within the device is due to laser power absorbed in the gold FSS and $VO_2$ layers [domains 1 and 3 in Fig. 4(c)]. The laser beam used in our model had a uniform intensity distribution within a radius of 50 μm. We used a continuous gold film to represent the gold FSS in our thermal modeling, which is expected to be a valid assumption because the area density of gold in the FSS is high (~0.88) and the apertures are small. We assumed the thermal conductivity and specific heat capacity of $VO_2$ to be 6 W/(m·K) and 690 J/(K·kg), respectively[34,35]. Note that these values correspond to $VO_2$ in its insulating phase, but their change due to the IMT is not dramatic (less than 20% across the IMT)[35], so for simplicity we set them to be constant. Thermal properties of gold and GaAs were taken from refs. 36 and 37.

Our model works as follows to calculate the transmittance given a particular incident intensity for a given bias temperature. With no laser irradiation, the device is in thermal equilibrium at the given bias temperature. The transient thermal simulation is initiated by heat flux induced into the device by laser irradiation. For each increasement in time, the simulation returns a transient temperature distribution within the device, which is used to update the absorptance based on the measured temperature-dependent absorption in Fig. 4(a). The resulting absorptance is then fed into the simulation at the next time step. This coupled opto-thermal simulation loop iterates until the temperature distribution stabilizes. Finally, we can convert the stabilized temperature distribution to transmittance according to Fig. 4(b). Based on convergence tests, we chose to use a step of 0.1 µs and a total simulation time of 500 µs (see details in *Supplementary Information 6*).

Both the limiting threshold and the trend of reduction of transmittance with incident intensity are well captured in our simulations [Fig. 4(d)], and agree very well with experimental data extracted from Fig. 3(c) (see Supplementary Information 4 for more details). Our simulations also confirmed that our experiments never fully reached the complete metal phase of the $VO_2$ due to the limited incident intensities (< 6.5 kW/cm$^2$). One can expect that once the IMT is completed, the output power will start to increase again with a very slow rate of ~0.03 (i.e., the limiting-state transmittance) as the input power increases.

Our opto-thermal modeling can be used to estimate the response time of the limiter (i.e., the time needed to reduce transmittance by $1/e$) by converting the temperature distribution at each time step to the transmittance. The time evolution of the transmittance for different incident intensity and bias temperatures is shown in Fig. 4(e, f). For example, given an incident intensity of 20 kW/cm$^2$, the response time is ~20 µs. The response time reduces to < 2 µs when the incident intensity is greater than 90 kW/cm$^2$. Similarly, for a given incident intensity (e.g. 90 kW/cm$^2$) one can expect to have a faster response at a higher bias temperature [Fig. 4(f)]. Note that the recovery time (from the limiting state to the open state, after the light turns off) is expected to be much shorter than the limiting response time, but the recovery time is not particularly important for limiting applications.

## Discussion

Both our experimental and simulation results indicate that one can engineer the limiting threshold by tuning the bias temperature. However, passive devices (i.e., without thermal bias) are more practical. In the absence of thermal bias, the limiting threshold can be engineered via a reduction of the IMT temperature using doping[38] or defect-engineering[39]. On the other hand, to increase the limiting threshold, one can reduce the open-state absorption by redesigning the FSS[40].

It is also important to consider the effect of the hysteresis in $VO_2$[41]. For most applications, it is favorable that the limiter automatically reverts to the open state once the incident light shuts off. Therefore, the device should not be thermally biased to any temperature within the hysteresis loop [e.g., temperatures greater than 52 °C according to Fig. S3(c) in *Supplementary Information*]. In our experiments, we did not demonstrate nonlinear transmittance with bias temperatures below 52 °C due to the limited output power of our laser. However, our simulations predict that this can be achieved when the incident intensity surpasses ~15 kW/cm$^2$ [the purple curve in Fig. 4(d)]. Alternatively, one can also consider exploring phase transitions with minimal hysteresis (e.g. the IMT in samarium nickelate[42]).

Finally, we note that in our experimental demonstration, the IMT in VO$_2$ was photothermally triggered by a continuous-wave laser. Therefore, the device speed is limited to ~microseconds. However, the IMT in VO$_2$ can actually be triggered non-thermally on time scales as short as tens of femtoseconds using high-intensity optical pulses[43], and therefore we anticipate that VO$_2$-based limiters can also limit very-high-intensity pulses at these time scales. The optimization of VO$_2$-based limiters for ultrafast operation may require a re-design, optimizing for higher electric-field concentration within the VO$_2$ layer to more-easily trigger the IMT [44].

To conclude, reflective optical limiters, which are transmissive at low input powers but reflecting at high powers, are a promising technology for sensor protection, because they can avoid damage due to light absorption in the limiter itself. Here, we explored the use of resonant transmission through metallic frequency-selective surfaces (FSSs) as the basis for reflective limiters, using the insulator-to-metal transition (IMT) in vanadium dioxide (VO$_2$) to make the resonant transmittance sensitive to incident intensity. The IMT in VO$_2$ can be a source of giant photothermal nonlinearity, which enabled the use of low-quality-factor FSS resonances in our design, resulting in open-state transmittance with large angular and temporal bandwidth. Our prototype reflective limiter designed for a wavelength of 10.6 μm has high open-state transmittance (~0.7), low limiting-state transmittance (< 0.01) and absorptance (~0.06), broad working bandwidth (FWHM > 2 μm), and functions for all polarizations up to an incident angle of ~50°.

## Materials and methods

VO$_2$ was deposited onto double-side-polished undoped GaAs (001) wafers via magnetron sputtering from a V$_2$O$_5$ target, with radio-frequency power of 100 W. During deposition, the chamber pressure was maintained at 5 mTorr with an Ar/O$_2$ gas mixture at a flow rate of 49.85/0.15 sccm. The substrate was heated to 700 °C to form the VO$_2$ phase.

We built a 50-nm gold FSS on top of the as-grown VO$_2$ via the following steps. First, a ~250-nm PMMA (495 PMMA A4) was spin-coated onto the VO$_2$ film and the pattern was written using an e-beam lithography system (Elionix GS-100). After development in MIBK/IPA (volume ratio of 1/3), an array of cross-shape PMMA blocks were left on top of the VO$_2$ film. Then, 50 nm of gold was evaporated, and the sample was soaked in an acetone bath for a few minutes to lift off the gold-PMMA blocks, leaving an array of cross-slit apertures within the gold film. Note that a 60-second sonication was necessary during the lift-off process.

## Acknowledgements

MK is supported by the Office of Naval Research (ONR, N00014-16-1-2556 and N00014-20-1-2297), with partial support from the Air Force Office of Scientific Research (AFOSR, FA9550-18-1-0146). SR is supported by the AFOSR (FA9550-18-1-0250).

## Conflict of interests

The authors declare no conflict of interest.

## Contributions

C.W. and M.K. conceived of the project and designed the experiments and simulations. C.W., J.K., Y.X., and R.W. performed the simulations. Z.Z. synthesized the $VO_2$ films. C.W. and Z.Y. conducted the fabrication. C.W., J.S., and A.S. built the experimental setup and carried out the experiments. All authors discussed the results and contributed to the writing of the manuscript. M.K. and S.R. supervised the project.

# Supplementary Information:
# Ultrathin broadband reflective optical limiter

Chenghao Wan[1,2], Zhen Zhang[3], Jad Salman[1], Jonathan King[1], Yuzhe Xiao[1], Zhaoning Yu[1,4], Alireza Shahsafi[1], Raymond Wambold[1], Shriram Ramanathan[3], and Mikhail A. Kats[1,2,4]*

[1] *Department of Electrical and Computer Engineering, University of Wisconsin-Madison, Madison, Wisconsin 53706, USA*
[2] *Department of Materials Science & Engineering, University of Wisconsin-Madison, Madison, Wisconsin 53706, USA*
[3] *School of Materials Engineering, Purdue University, West Lafayette, IN 47907, USA*
[4] *Department of Physics, University of Wisconsin-Madison, Madison, Wisconsin 53706, USA*

*Email: mkats@wisc.edu*

## Section 1. Design and optimization of the FSS-VO$_2$ limiter

For the choice of the antenna geometry, we first made sure to select a highly symmetric structure that provides an isotropic response (i.e., no polarization dependence at normal incidence). There are many candidates, including square apertures [Fig. S1(a)], coaxial square apertures [Fig. S1(b)], cross-slit apertures that we used [Fig. S1(c)], etc.[S1,S2,S3]. For each candidate structure, we can achieve resonance transmission for a desired wavelength [e.g., ~7 μm shown in Fig. S1(a – c)] by adjusting the size and periodicity of the apertures. The different line shapes of the resonant transmission are attributed to multiple factors, including aperture geometry, structure symmetry, and the dielectric environment (i.e., the thicknesses and dielectric constants of VO$_2$ and GaAs). Comprehensive interpretation of these factors can be found in refs. [S2, S3]. We found that, among the shapes we explored, the high-density array of cross-slit apertures can feature the broadest transmission bandwidth when VO$_2$ in the insulating phase [Fig. S1(d)], which is favorable for our application. Empirically, we also found that although a larger density of the aperture antennas increases the bandwidth and amplitude of the open-state transmission, it results in a higher absorptance in the limiting state [Fig. S1(d)].

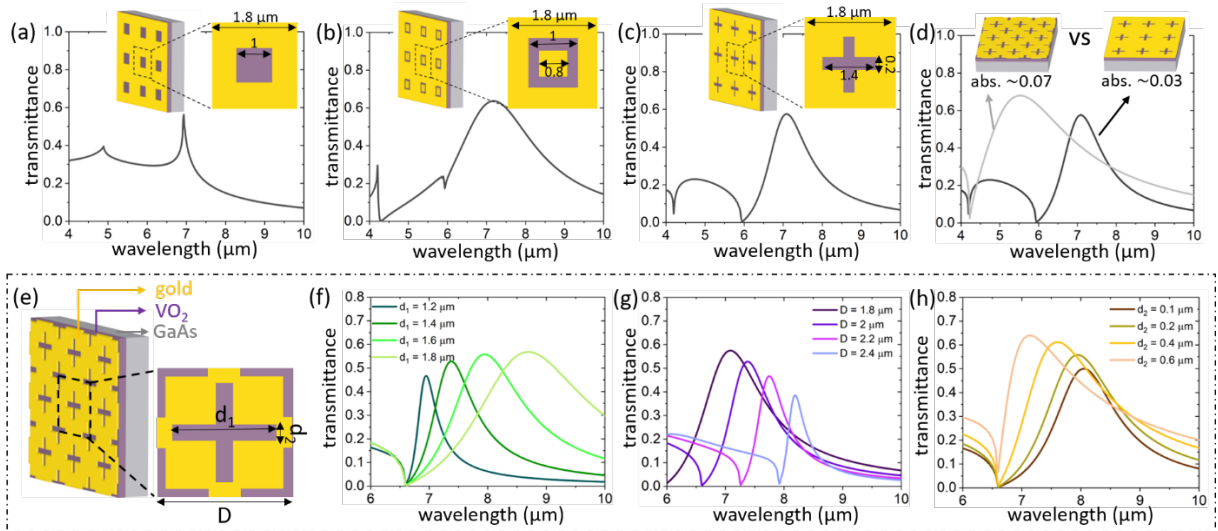

**Figure S1. (a)** Simulated open-state transmittance of the FSS-VO$_2$ limiter with **(a)** square-shaped apertures **(b)** coaxial square apertures, and **(c)** cross-slit apertures. **(d)** Simulated open-state transmittance of the FSS-VO$_2$ limiter with a high-density (grey curve) and a low-density (black curve) array of aperture antennas. As shown in the insets, the high-density array comprises close-packed apertures as those of Fig. 1(b) in the main text, setting $d_1$ = 1.4 μm, $d_2$ = 0.2 μm, and D = 1.8 μm. The low-density array is generated from the high-density one by removing every other row of antennas. The corresponding limiting-state absorptances are given in the insets. **(e)** A schematic of the FSS-VO$_2$ limiter that we eventually chose for demonstration, which is the same as that of Fig. 1(a) in the main text. **(f)** Simulated open-state transmittance of the FSS-VO$_2$ limiter for $d_1$ = 1.2, 1.4, 1.6, and 1.8 μm, with fixed $d_2$ = 0.2 μm and D = 2 μm. **(g)** Simulated open-state transmittance of the FSS-VO$_2$ limiter for D = 1.8, 2, 2.2, and 2.4 μm, with fixed $d_2$ = 0.2 µm and $d_1$ = 1.6 µm. **(h)** Simulated open-state transmittance of the FSS-VO$_2$ limiter for $d_2$ = 0.1, 0.2, 0.4, and 0.6 µm, with fixed D = 2 µm and $d_1$ = 1.6 µm.

The optimization parameters of our design [Fig. S1(e)] include the thicknesses of the gold and VO$_2$ layers, the length ($d_1$) and width ($d_2$) of the aperture antennas, and the periodicity (D). Compared to the other parameters, the thicknesses of the gold and VO$_2$ layers have much less influence on tuning the spectral location of the resonance peak of the FSS, so long as the thicknesses are in the deep-subwavelength range (thickness of ~100 nm ≪ operational wavelength of 10.6 µm). Therefore, we fixed the thicknesses of gold and VO$_2$ to be 50 nm and 100 nm, respectively. Similarly, the primary parameters that determine the resonance frequency, and therefore the peak wavelength of the open-state transmittance, are the aperture length, $d_1$ [Fig. S1(f)] and the periodicity, D [Fig. S1(g)]. The quality factor (Q) of the FSS is closely related to the area density of the aperture antennas, which is determined by both D and $d_2$. As shown in Fig. S1(g, h), D and $d_2$ are coupled, and both influence the amplitudes and bandwidth of the open-state transmittance peak. Note that $d_2$ and D also affect the spectral position of the open-state transmittance peak, although not as significantly as $d_1$ does.

## Section 2. Bare VO$_2$ film vs. FSS-VO$_2$ limiter

Using finite-difference time-domain (FDTD) simulations, we compared the open- and limiting-state transmittance and absorptance spectra of a bare VO$_2$ film on a GaAs substrate to those of the FSS-VO$_2$ limiter shown in Fig. 1(b) in the main text. For a bare 100-nm VO$_2$ film on GaAs, the open-state transmittance is

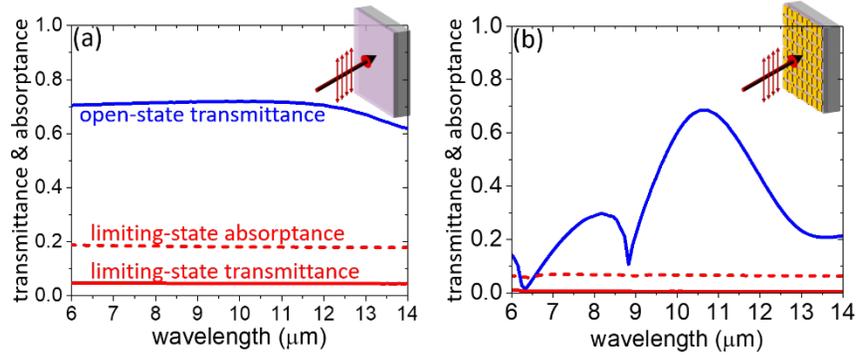

**Figure S2.** Simulated transmittance and absorptance spectra for **(a)** a bare 100-nm VO$_2$ film on a GaAs substrate and **(b)** the FSS-VO$_2$ limiter shown in Fig.1(b) in the main text.

~0.7 across a broad spectral regime. The limiting-state transmittance is ~0.05 [Fig. S2(a)]. The bare film has a limiting-state absorptance of ~0.2. With the FSS on top, the limiting-state transmittance is reduced to < 0.01, and the limiting-state absorptance is reduced to ~0.06, which significantly improves the device damage threshold. Note that compared to limiters that are merely based on thin-film VO$_2$ (e.g., ref. S3), incorporating resonant structures results in reduction of other figures of merit such as open-state transmittance, spectral and angular bandwidths. However, due to the ultra-high photothermal nonlinearity of VO$_2$, the quality factor of the FSS can be quite small [FWHM > 2 μm, as shown in Fig. S2(b)], thus minimizing these compromises.

## Section 3. Post-growth characterization of VO$_2$ film

The thickness of the resulting film is ~105 nm, measured by scanning electron microscopy (SEM, Zeiss LEO 1530) imaging of the cross section [Fig. S3(a)], which is close to our target thickness (100 nm) that was used in our simulation. Atomic force microscopy (AFM, Bruker MultiMode 8) imaging confirmed that the film was continuous with a surface roughness of Ra ≅ 6 nm [inset of Fig. S3(a)]. The stoichiometry of the film was confirmed by Raman spectroscopy measurements (LabRAM ARAMIS, Horiba) at 30 °C and 100 °C, with the pump laser operating at 520 nm [Fig. S3(b)]. At 30 °C, signature Raman modes of insulator-phase VO$_2$ at 140, 192, 223, 308, 387, 395, 613, and 823 cm$^{-1}$ were detected [S5]; while at 100 °C, the VO$_2$ is in the metallic state and only features of the GaAs substrate were observed. We also note that we did not find any clear Raman modes of V$_2$O$_5$ to confirm that there is a surface oxide (i.e., V$_2$O$_5$) on top of the as-grown VO$_2$ film, as indicated in Fig. 2(e) in the main text. One plausible explanation is that such surface oxide may be too thin (and perhaps not continuous) to be identified by our Raman measurements.

In our previous study (ref. [22] in the main text), we also found that the appearance of such surface oxide [i.e., the feature at ~10 μm in Fig. 2(e)] is dependent on substrate choice and synthesis method; more discussion can be found in that paper.

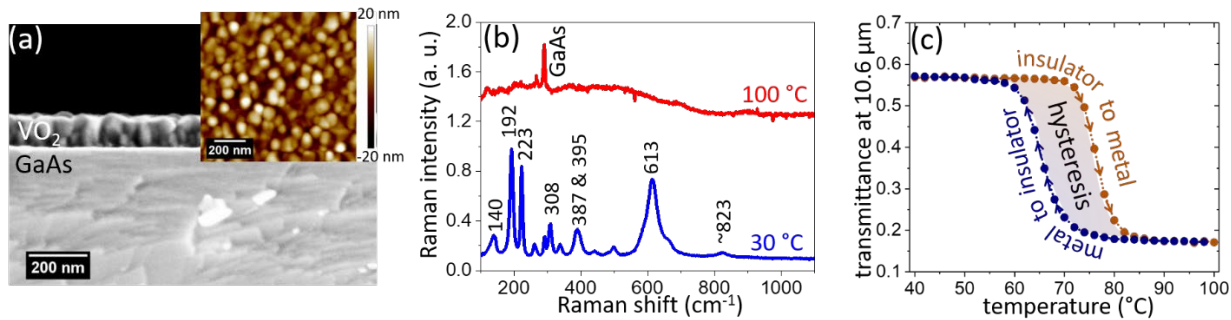

**Figure S3. (a)** Cross-section SEM of the as-grown $VO_2$ on GaAs. The inset is AFM characterization of the film surface, indicating the film roughness of Ra ~6 nm. **(b)** Raman spectra of the as-grown $VO_2$ at 30 and 100 °C. At 30 °C, signature modes of insulator-phase $VO_2$ were identified, while at 100 °C only peaks of the GaAs substrate were observed. **(c)** Temperature-dependent transmittance at λ = 10.6 μm, measured under heating (the brown curve) and cooling (the navy-blue curve) processes.

The IMT of the film was confirmed by temperature-dependent near-normal-incidence transmission measurements using a Fourier-transform spectrometer (FTS, Bruker Vertex 70) connected to an infrared microscope (Bruker Hyperion 2000) [Fig. 2(b) in the main text]. All spectra were collected at temperatures between 30 °C and 100 °C (first heating, then cooling) with steps of 2 °C and a ramping rate of 1 °C/min. The as-grown film featured an insulator-to-metal transition from ~70 to 82 °C when heated and a metal-to-insulator transition from ~72 to 54 °C when cooled [Fig. S3(c)].

**Section 4. FDTD simulations vs. experimental results of the fabricated limiter**

We found discrepancies in the amplitudes and peak positions of the open-state transmittance when comparing the measured spectra of the fabricated FSS-$VO_2$ limiter [Fig. 2(f)] to our simulations [Fig. 1(b)]. The amplitude of the experimental open-state transmittance peak is lower than that of the simulation, likely because we did not consider the backside of the substrate in the FDTD simulation. After taking it into account, the simulated peak value matches well with that of the experimental result [Fig. S4(b)]. Also, the open-state central wavelength of the fabricated limiter is blue-shifted compared to our simulation results. We believe that the primary reason for the blue shift is that the aperture length and width in the fabricated FSS are ~3 µm and ~0.17 µm, respectively [Fig. S4(a)], both smaller than those we used in simulations, where $d_1$ = 3.1 µm, $d_2$ = 0.2 µm. Another possible contribution to this blue shift is that the refractive indices of $VO_2$ that we used for simulation were not extracted from the film that we used for fabrication. Some differences in refractive-index values are expected between $VO_2$ films, though they should be small in this wavelength region[S4].

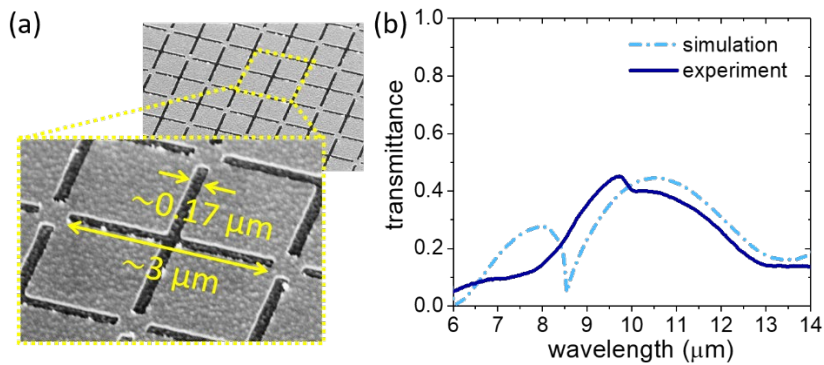

**Figure S4. (a)** SEM image of the fabricated FSS-$VO_2$ limiter. **(b)** Comparison of open-state transmittance between the experiment in Fig. 2(f) and simulation in Fig. 1(b) with the backside of GaAs included and the dimensions matching the SEM in "(a)".

## Section 5. The incident intensity in the experiments

We assumed the intensity distribution within the beam spot followed a Gaussian function. The beam radius ($w_0$) is defined to be the distance from the beam axis where the intensity drops to $1/e^2$, and can be approximated using Eq. S1, where $\sigma$ is the standard deviation of the Gaussian [Fig. S5(a)][S7]. To characterize the size of the focal spot, we evaporated a 50-nm-thick, 250-µm-wide gold bar onto a double-side-polished GaAs wafer [the inset of Fig. S5(b)]. When the gold bar was placed at the focus and centered with respect to the the axis of the beam [Fig. S5(b)], the power transmitted through the sample ($P_{block}$) was minimized. When the gold bar was moved to far from the beam, we recorded the maximum transmission ($P_{no\ block}$, i.e., the transmission of the double-polished GaAs wafer). We estimated the radius of the focal spot using Eqs. S1 – S3, where $\Phi(z)$ is the cumulative distribution function of the normal standard distribution[S7]. Then we calculated the peak intensity (i.e., the central intensity at the focus) using Eq. S4[S6].

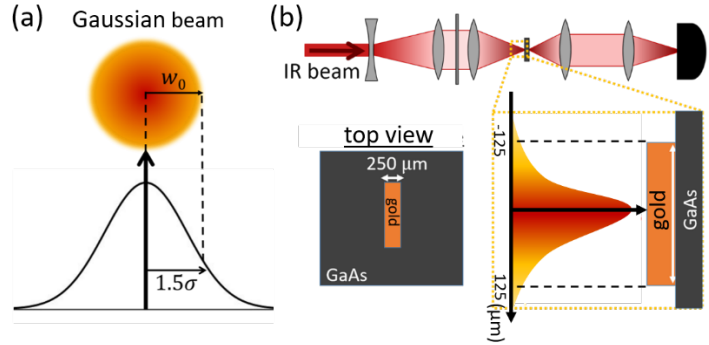

**Figure S5. (a)** The radius ($w_0$) of a Gaussian beam is defined to be 1.5-fold of the standard deviation (σ) of the power-distribution function. **(b)** Our method of characterizing the beam radius at the focus, using a 50-nm thick, 250-µm-wide gold bar evaporated on a GaAs wafer and mounted on a translation stage.

$$w_0 = 1.5\ \sigma \qquad \text{(Eq. S1)}$$

$$P_{block}/P_{no\ block} = 2(1 - \Phi(z)) \qquad \text{(Eq. S2)}$$

$$z = (width\ of\ gold\ bar)/(2\sigma) \qquad \text{(Eq. S3)}$$

$$\text{Peak intensity} = (2 \cdot power)/(\pi w_0^2) \qquad \text{(Eq. S4)}$$

**Section 6. Transient simulation using COMSOL Multiphysics**

We built the mesh for all domains using the pre-defined "extra-fine" option in COMSOL Multiphysics [Fig. S6(a)]. With this fine meshing, the time step needs to be sufficiently small to capture the dynamic temperature evolution that is caused by the laser irradiation. We performed simulations with time steps of 0.5, 0.1, and 0.05 µs, given an incident intensity of 25 kW/cm² and a bias temperature of 52 °C. As shown in Fig. S6(b), once the device is photothermally heated to the onset of the IMT regime (between ~70 and 76 °C), there are temperature fluctuations that are caused by the decrease of the absorptance starting at ~74 °C [Fig. 4(a) in the main text]. These dynamics in temperature were well captured and converged when the time step was set to < 0.1 µs.

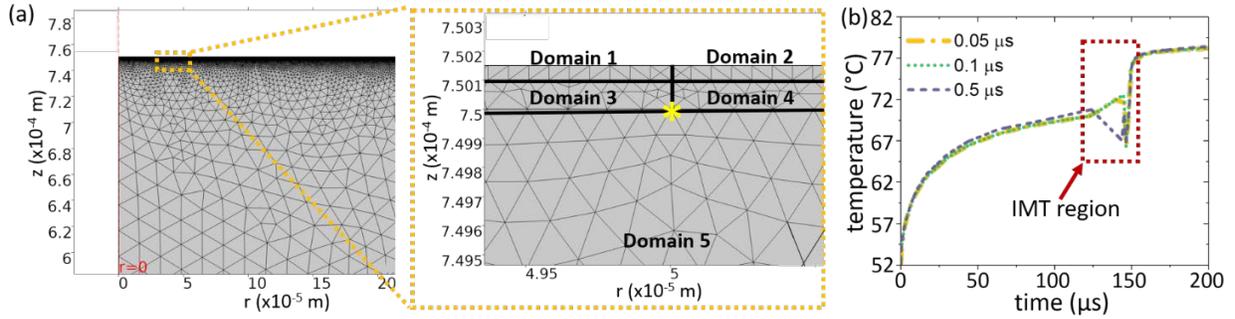

**Figure S6. (a)** Mesh setup for all domains using the "extra-fine" option in COMSOL Multiphysics **(b)** Convergence test by performing simulations with time steps of 0.5, 0.1, and 0.05 µs. The temperatures were extracted from the point at r = 50 µm, z = 750 µm, which is at the edge of the beam spot, as denoted by the asterisk in "(a)".